\documentstyle[12pt,epsfig]{article}
\def\gsim{\ \rlap{\raise 2pt \hbox{$>$}}{\lower 2pt \hbox{$\sim$}}\ }
\def\lsim{\ \rlap{\raise 2pt \hbox{$<$}}{\lower 2pt \hbox{$\sim$}}\ }
\newskip\humongous \humongous=0pt plus 1000pt minus 1000pt

\newif\ifdtup

\begin{document}
\centerline{\bf MIXED DARK MATTER FROM AXINO DISTRIBUTION}

\medskip\medskip\medskip

\rm

\centerline{{\bf S.A.~Bonometto}$^{1,2}$, {\bf F.~Gabbiani}$^3$ and
{\bf A.~Masiero}$^4$}

\medskip\medskip\medskip

\rm
\noindent
$^{1}$ Dipartimento di Fisica dell'Universit\'a di Milano,
Via Celoria 16, I-20133 Milano, Italy

\noindent
$^{2}$ I.N.F.N. -- Sezione di Milano

\noindent
$^{3}$ Dept. of Physics and Astronomy, University of Massachusetts
at Amherst, Amherst, MA 01003, USA

\noindent
$^{4}$ I.N.F.N. -- Sezione di Padova, Via Marzolo 8, I-35131 Padova,
Italy

\vglue 2.8truecm
\centerline{\bf Abstract}
\bigskip

We study the possibility of mixed dark matter obtained through the phase
space distribution of a single particle. An example is offered in the
context of SUSY models with a Peccei-Quinn symmetry. Axinos in the 100
keV range can naturally have both thermal and non-thermal components.
The latter one arises from the lightest neutralino decays and
derelativizes at $z \sim 10^4$.

\vglue 1.2truecm

PACS number(s): 95.35.+d, 98.80.--k

\vfill\eject

\section{Introduction}

\noindent
Large scale structure data favour hybrid cosmological models. That dark matter
(DM) consists of different components is an idea which dates back to the middle
eighties, when the linear stages of hybrid models were fully
evaluated$^{(1)}$.
The failure of canonical cold dark matter (CDM) models recently renewed the
interest on such models. A recent evaluation of their non-linear
stages$^{(2)}$ shows that it eases the solution of the problems CDM
could not solve. Recent work$^{(3)}$ is based on a mixture of massive
neutrinos with $m_{\nu}$ $\simeq$ 7 eV (hot dark matter: HDM) and
other non-baryonic particles yielding CDM. The HDM density parameter
$\Omega_{\nu}$ = 0.3($m_{\nu}$/28$\, $h$^2$eV) (h: Hubble parameter
in units of 100km$\, $s$^{-1}$ Mpc$^{-1}$, in this note we take $h^2$ 
= 0.5). Baryonic matter, expected to have a
marginal role in the origin of inhomogeneities, has a density parameter
$\Omega_{\rm B}$ $\lsim$ 0.05, because of primeval nucleosinthesis
constraints. The cold component is then assumed to have a density
parameter $\Omega_{\rm CDM}$ $\simeq$ 0.6, in order that $\Omega$ = 1.
Massive neutrinos derelativize at a redshift $z_{\rm der} \simeq
10^4$($m_{\nu}$/7eV).

With respect to canonical CDM, the above CHDM hybrid model has only
one extra parameter ($m_{\nu}$). Nevertheless it is hard to escape
the feeling that having $\Omega_{\rm HDM}/\Omega_{\rm CDM}$ of
{\cal O}(1) is a fortuitous coincidence. In this note we propose a
hybrid model where a single particle naturally accomplishes the twofold
role of CDM and HDM, thanks to its expected phase space distribution.
The value of the $\Omega_{\rm HDM}/\Omega_{\rm CDM}$ ratio is linked
to precise microphysical parameters. For a significant sector of the
parameter space, we expect our model to reproduce (or to improve)
CHDM outputs. It is clear that the {\sl quality} of a model cannot
be strictly linked to the request of minimizing the total parameter
amount. Our choice of microphysical parameters is fairly natural
and testable at present or forseeable experimental facilities.

Our framework is based on the supersymmetric (SUSY) extension of
the standard model where a Peccei-Quinn (PQ) symmetry$^{(4)}$ is
implemented to solve the strong {\cal CP} problem. More
specifically, we consider a SUSY version of the invisible DFS
axion$^{(5)}$. Accordingly, beside
axions, the model predicts the existence of its fermionic partner
called axino ($\tilde a$). 

In order to obtain MDM, such
framework is to be restricted as follows:
(i) The lightest neutralino ($\chi$)
is an (almost) pure gaugino. (ii)
The SUSY soft breaking scale, related
to the sfermion ($\tilde f$) masses is of {\cal O}(TeV). 
(iii) The PQ scale ($V_{\rm PQ}$) is in the lower side of the interval allowed
by astrophysical constraints$^{(6)}$, ranging about 10$^{10}$GeV.

As the axino mass $m_{\tilde a}$
$\sim$ $m^2_{\tilde f}/V_{\rm PQ}$, taking $V_{\rm PQ}$
$\sim$ 10$^{10}$ GeV and $m_{\tilde f}$
$\sim\, $ TeV, leads to $m_{\tilde a}$ of {\cal O}(100 keV).
In turn this would be coherent only with a present CBR temperature
$T_o$ $\ll$ 2.7 K (or, if we require $T_o$ = 2.7 K, with
$\Omega$ $\gg$ 1, well above observational limits) unless a
substantial entropy release took place between axino decoupling and
today, {\it e.g.} at the electroweak phase transition$^{(7)}$. Such
release rises the CBR temperature at the observed level.

In two previous works it was already suggested that DM consists
of $\tilde a$'s. In one of them$^{(8)}$ it was assumed that DM
$\tilde a$'s originated because of $\chi$ decay. In the other
one$^{(9)}$ it was shown that this could hardly account for
$\Omega$ = 1, while $\tilde a$'s, formerly in thermal equilibrium
with the other components of the Universe and decoupling at a
temperature $T_{{\tilde a},{dg}} < V_{\rm PQ}$, can easily
account for $\Omega$ = 1. In this note we take into account both
thermal and non-thermal axino components. The former component
will be an effective CDM, as only fluctuations involving masses
M $\lsim$ 0.1 M$_\odot$ will be erased at its derelativization.
The latter component, instead, can easily behave as HDM,
derelativizing at a redshift $z \sim$ 10$^4$. Accordingly,
fluctuations in such component will be erased up to a mass 
M$_{\rm D}$ $\sim$ 10$^{15}$ M$_{\odot}$. The
exact values of the ratio $\Omega_{\rm HDM}/\Omega_{\rm CDM}$
and of $z_{\rm der}$ depend on $m_{\chi}$ (neutralino mass),
$m_t$ (top quark mass), and $m_{\tilde f}$.

\section{Non-thermal Axino Production}

\noindent
Axinos are the natural lightest SUSY fermions; therefore $\chi$'s
can decay, according to the reaction $\chi \rightarrow \tilde a
+ \gamma$ (we assume R-parity to be conserved). The temperature
of the Universe when this occurs is given by:
$$
T^4_{dy} = 45 m^2_{pl} \Gamma^2_{dy} / 4\pi^3 g_{\chi,dy}.
\eqno(1)
$$
Here $m_{pl}$ = $G^{-1/2}$ = 1.2$\times 10^{19}$ GeV, $g_{\chi,dy}$
= ${\cal N}_b+(7/8){\cal N}_f$ [${\cal N}_{b(f)}$ is the number of
independent spin states of bosons (fermions) of particles with
$m < T_{dy}$, still coupled at $T_{dy}$], and
$$
\Gamma_{dy} = (\alpha^2_{\rm em}/18 \pi^3)(m^3_{\chi}/
V^2_{\rm PQ})\vert f(y) \vert^2
\eqno(2)
$$
is the decay rate$^{(10)}$, which is equated to the cosmic expansion rate.
In eq. (2) $f(y) = y(1-y+\ln y)/(1-y)^2$ and y = $m^2_t/m^2_{\tilde f}$
(typical $|f(y)|$ $\sim$ 10$^{-2}$). When the decay occurs, the
$\chi$'s have
decoupled since long. This occurred at a temperature $T_{\chi,dg}$
when $x$ = $T_{\chi,dg}/m_{\chi}$ was already below unity: $x$ is
obtained equating the cosmic expansion rate with the rate of the
reaction $\chi\chi \rightarrow f\overline{f}$. Assuming that
$\chi$ is a pure gaugino is crucial to restrict the dynamics of the
reaction to pure sfermion exchange (we consider Higgs exchange
contribution to be negligible). According to ref. (11) we obtain
$$
e^{1/x}x^{-3/2} = (8\alpha_{\rm em}^2/\pi^2)(45/2g_{\chi,dg})^{1/2}N
m^3_{\chi}m_{pl}/m^4_{\tilde f},
\eqno(3)
$$
with $N$ = 116/27 = $\sum_f q^4_f n_{c,f}$ ($q_f$: fermion charge; $n_c$: color
number), while $g_{\chi,dg}$ is the analogue of $g_{\chi,dy}$, but at $T_{dg}$
instead of $T_{dy}$. The number density of residual $\chi$'s is then 
$$
n_{\chi,dg} = 0.70n_{\rm rel}(T_{\chi,dg})x^{-3/2}e^{-1/x},
\eqno(4)
$$
where $n_{\rm rel}(T)$ = $(\zeta(3)/\pi^2)g_nT^3$, while $g_n$ =
${\cal N}_b+(3/4){\cal N}_f$ (therefore, in our case,
$g_n = 1.5$). The number of $\tilde a$, arising
from $\chi$ decay, at any $T \ll T_{\chi,dy}$, is $n_{\tilde a}$
= $n_{\chi,dg}[a_{\chi,dg}/a(T)]^3$, $a$ being the scale factor,
while their linear momentum, in average, is 
$(m_{\chi}/2)[a_{\chi,dy}/a(T)]$. 
The
detailed momentum distribution is quite far from thermal.
It can be easily obtained taking into account that each $\tilde a$,
when it arises (from non--relativistic $\chi$'s), has momentum
$p = m_\chi/2$. Let $t^*$ be the time 
when a given $\tilde a$ is produced, the comoving
number of $\chi$'s then reads
$$
n_\chi ~=~ n_{\chi,dg} \exp(-2 \Gamma_{dy} t^*) ~.
\eqno(5)
$$
At any time $t \ll \Gamma_{dy}^{-1}$, the momentum distribution of
$\tilde a$'s is obtainable from the relation
$$
{d n_{\tilde a} \over d p} ~=~ -2 {t^* \over z^*}
{d n_{\tilde a} \over d t^*} {d z^* \over d p} 
\eqno(6)
$$
(redshift and scale factor are related according to $1+z$ = $a_o/a$).
Here $p = m_\chi z/2\, z^*$ is the momentum of $\tilde a$'s
at the time $t$; $z$ and $z^*$ are the redshifts at the times
$t$ and $t^*$ respectively. Owing to eqs. (5) and (6), the
distribution of $\tilde a$'s on momentum, at the redshift $z$, reads
$$
\Phi_z(p) ~=~ {1 \over n_{\chi,dg}} {d n_{\tilde a} \over d p} 
~=~ 2 {q \over p} e^{-q}
\eqno(7)
$$
with $q = [(z_{\chi,dg}/z) (2p/m_\chi)]^2$.
For $p$ = $m_{\tilde a}$ axinos derelativize. In average, this takes
place at a redshift given by
$$
z_{\rm der}/10^4 \simeq 0.2|f|(m_{\tilde a}/keV)(m_{\chi}/
10~{\rm GeV})^{1/2} (V_{\rm PQ}/10^{10}~{\rm GeV})^{-1}.
\eqno(8)
$$
Eq.~(8) immediately shows that, taking $m_\chi$ and $m_{t}$ (from which the
value of $f$ follows) in the experimentally admitted range, and choosing low
$V_{PQ}$ and high $m_{\tilde f}$ (from which the value of $m_{\tilde a}$
follows), $z_{\rm der}/10^4 \simeq 1 $ is a generic consequence. Quite
independently of their mass and because of their non--thermal distribution,
axinos derelativize at the same epoch as massive neutrinos with mass $\sim 7\,
$eV, and a previously thermal distribution. 

Together with the right derelativization, axinos also have a fair number
density. Let us stress that $n_{\chi}$ and, therefore, $n_{\tilde a}$ have a
strong dependence on $m_{\tilde f}$. This can be seen through eqs. (3) and (4).
If powers of $x$ are neglected with respect to exponentials, it turns out that
$n_{\tilde a} \propto m_{\tilde f}^4$. With the large values of $m_{\tilde f}$
considered here, $\chi$'s in the allowed mass range would close the Universe,
unless they decay into $\tilde a$ (or other particles with masses $\lsim
10^3$keV).  

Let us also notice that, once we require that $m_\chi/m_{\tilde a} \gsim
10^5$, having $z_{der} \sim 10^4$ leads to $z_{\chi,dy} \gsim 10^9$. However,
the value of  $z_{\chi,dy} $ is soon obtainable from eq.~(2). The point is
that, within the above range of $V_{PQ}$ and $m_{\tilde f}$, the right value of
$\Gamma_{dy}$ follows from taking the top mass in the 150$\, $GeV range,
according to experimental constraints. 

Decay photons, emitted at $z \sim 10^9$ ($T \sim 230\, $keV) are thermalised
and cause no observable distortion of CBR spectrum. No appreciable amount of
nuclides, produced in big--bang nucleosynthesis, is likely to be destroyed by
decay photons. We shall devote the rest of this section to show this point. 
The essential argument is that, at least down to $T \sim 20\, $keV,
nuclides are screened against $\chi$--decay photon disruption either by CBR
photons or by the residual thermal electrons. Then, at $T < 20\, $keV,
the residual number of $\chi$ is however too small to cause any appreciable
consequence.

At any given temperature $T < 230\, $keV, the residual number density of $\chi$
reads 
$$
n_\chi (T) \sim 10^{-5} n_\gamma e^{-(230\, {\rm keV}/ T)^2} ~.
\eqno (9)
$$
Here $n_\gamma$ is the thermal photon number density at the same $T$. (If
$m_\chi$ exceeds 30 GeV, $n_\chi $ would be even smaller.) Photons arising from
$\chi$ decay are well above threshold for the production of $e^+,e^-$ pairs in
the collision with CBR. At $T \sim 100\, $keV the cross--section for this
process is $\sim 10^{-3} \sigma_T$ ($\sigma_T$: Thomson cross section), while
the cross--section for the reaction $\gamma + ^2$H$ \to p + n$ is $\sim 10^{-6}
\sigma_T$. $^2$H disruption by high energy $\gamma$'s is therefore a negligible
process. 

However, this is not enough to show that nuclide disruption is unlikely.
In fact, in the $e^+,e^-$ pair production most energy goes to a single
electron. Such electron, via inverse--compton in the Klein--Nishina regime, is
going to yield almost all its energy to another photon, etc. This creates a
cascade process$^{(12)}$ and the low energy by--products, both electrons and
$\gamma$'s, are potentially dangerous to $^2$H, which is no longer shielded by
CBR $\gamma's$, as low energy by--products are mostly below threshold for 
$e^+,e^-$ pair production. 

As a matter of fact, at $T < m_e$, the thermal electron abundance is depressed
by a Boltzmann factor $\sim y^{3/2} e^{-y}$ with $y = m_e/T$, but, at $T \sim
100\, $keV, a considerable amount of electrons are still present. Only when $T
\sim 20\, $keV ($y \sim 25$) does the electron number density become so low, to
approach the baryon number density. Let us however recall that the cross
section for nuclide disruption is $\sim 10^{-6} \sigma_T$. The 
number of cascade by--products per $\chi$-decay is smaller than 10$^6$
and therefore residual electrons are a good shield to  $^2$H and other
nuclides. At $T \sim 20\, $keV, the residual $n_\chi \sim 10^{-48} n_\gamma$.
This compares with the small abundances of light nuclides produced in big--bang
nucleosynthesis. $E.g.$, the predicted $^7 Li$ abundance corresponds to a
number density $n_{[^7 Li]} \sim 10^{-18} n_\gamma$ and cannot be appreciably
modified by any photon cascade caused by $\chi$ decay. 

\section{Thermal Axino Production}

\noindent
Aside of non-thermal $\tilde a$'s, thermal $\tilde a$'s, decoupling
when the cosmic temperature $T_{\tilde a,dg}$ was slightly below
$V_{\rm PQ}$, are expected to exist$^{(9)}$. Their present number
density can be obtained from
$$
a^3_{\tilde a-dg}T^3_{\tilde a-dg} = a^3_{\tilde a-dg}[\pi^2/\zeta(3)
g_{n,\tilde a}]n_{\tilde a,{\tilde a}-dg} =a^3_o[\pi^2/\zeta(3)g_{n,\tilde a}]
n_{\tilde a,o}.
\eqno(10)
$$
($g_{n,\tilde a}$ = 3/2). After $\tilde a$ decoupling, the Universe underwent
(at least) the electroweak phase transition. If we consider the entropy
increase$^{(7)}$ at such transition(s), current mass limits for particles
decoupling earlier are modified. We shall take that into account by means of a
suitable $\mu$ factor in the following relation: 
$$
\mu(g_{\tilde a-dg} - g_{\tilde a})a^3_{\tilde a-dg}T^3_{\tilde a-dg}
= g_oa^3_oT^3_{\gamma,o} = (g_o/2)[\pi^2/\zeta(3)]a^3_on_{\gamma,o}.
\eqno(11)
$$
Here $g_o$ = $g_{\gamma}(g_{\nu} + g_{\gamma} + g_e)/(g_{\gamma} + g_e)$
= 43/11 ($n_{\gamma,o}$ is the present photon number density). Comparing
eqs. (10) and (11), after simple calculations, yields the present density
parameter $\Omega_{\rm CDM}$ of thermal $\tilde a$. This reads
$$
\Omega_{\rm CDM} = 0.51{ {m_{\tilde a}} \over {\rm keV}}{1\over {\mu h^2}}
{220 \over {(g_{{\tilde a}-dg} - g_{\tilde a})}}.
\eqno(12)
$$
The spin degrees of freedom
at $T_{\tilde a,dg}$ yield $g_{\tilde a,dg} \simeq$ 220. Both
non-thermal and thermal $\tilde a$'s are fully non-relativistic
today. Accordingly their present densities are $\rho_{o,{\tilde a}-
n.th}$ = $m_{\tilde a}n_{\chi,\chi-dg}(a_{\chi-dg}/a_o)^3$
and $\rho_{o,{\tilde a}-th}$ = $m_{\tilde a}n_{\tilde a,{\tilde a}-dg}
(a_{{\tilde a}-dg}/a_o)^3$. Then, owing to eqs. (4) and (12), we have that
$$
{ \Omega_{\rm HDM} \over \Omega_{\rm CDM} } \simeq
{ {0.70g_{\chi}(g_{{\tilde a}-dg} - g_{\tilde a})\mu} \over 
{g_{\tilde a}g_{\chi-dg}} }x^{-3/2}e^{-1/x}.
\eqno(13)
$$
In the above relations the microphysical variables are $m_{\tilde a}$,
$m_{\chi}$, $m_{\tilde f}$, $m_t$, $V_{\rm PQ}$. Other physical
variables are $\mu$, $\Omega_{\rm CDM}$, $\Omega_{\rm HDM}$ and
$z_{\rm der}$. Besides of eqs. (8), (11), (12), the constraint
$m_{\tilde a}V_{\rm PQ} \simeq m^2_{\tilde f}$ also holds.

Before concluding this section we wish to outline that the factor $\mu$
introduced here is also related to the baryon--to--photon ratio
after the electroweak transition (where baryon number could be
generated, if an out--of--equilibrium phase lasts long enough).
In principle this might allow to relate the final value of
the present baryon density parameter $\Omega_B$ to the final
ratio ${ \Omega_{\rm HDM} \over \Omega_{\rm CDM} }$.

\section{Conclusions}

\noindent
Within the frame of a SUSY implementation of the invisible DFS
axion approach, we can restrict the parameter space either considering the
expected ranges for $m_{\chi}$, $m_{\tilde f}$, $m_t$ and
$V_{\rm PQ}$, or by requiring that $z_{\rm der} \sim 10^4$,
$\Omega_{\rm CDM} \sim$ 0.6-0.8, $\Omega_{\rm CDM}/\Omega_{\rm HDM} \sim$
3-4. The point we wish to make is that there is a significant overlap
between the parts of the parameter space which are allowed by either
set of constraints. We illustrate this fact by means of figs. 1--2
where microphysical parameters are constrained to provide suitable
values of $z_{\rm der}$. $\Omega_{\rm CDM}$ = 0.8 and
$\Omega_{\rm HDM}$ = 0.2 are taken, but results do not depend
critically on this assumption.

Fig. 1 shows for which values of $m_{\tilde f}$ and $m_{\chi}$
we obtain $z_{\rm der} \simeq 10^4$. Different curves refer to the
top masses ranging from 120 GeV to 180 GeV. 
Fig. 2 shows the dependence of the $V_{\rm PQ}$ scale on
$m_{\chi}$, for $m_t$ = 140 GeV and requiring $z_{\rm der}
\simeq 10^4$. For $m_{\chi} \simeq$ 30 GeV, $V_{\rm PQ} \sim$
1.7$\times 10^{10}$ GeV. The corresponding value of  
$m_{\tilde a}$ is $\sim$ 180 keV, while $\mu \sim$ 220.
Fig. 3 is a description of the evolution of densities for different
components in the Universe, starting from $T \sim V_{\rm PQ}$ down
to today's temperature. This plot is meant to show
how complicate it can appear, {\sl a priori}, to
build up a component arising from heavier particle decay,
 with both the right derelativization
redshift ($z_{der}$) and final density parameter
($\Omega_{\rm HDM}$). The main issue of this paper is
that such aims are achieved fairly naturally in a SUSY implementation
of the DFS invisible axion approach, provided large SUSY scale and small
PQ scale are taken.

The detailed evolution
of density fluctuations over different scales, through
$\tilde a$ derelativization, equivalence,
hydrogen recombination should be computed in order to
evaluate CBR fluctuations, large scale structure and
velocity fields, galaxy mass function. A study of the
linear stages of this model will be presented elsewhere.
It is however possible to estimate the minimal mass 
of fluctuations able to survive derelativization both
for the cold and the hot parts of $\tilde a$ distribution
($M_{D,c}$ and $M_{D,h}$, respectively), just by evaluating the mass scales
entering the horizon at the time $t$ of derelativization.

For the cold component, 
$M_{D,c}$ $\simeq$ $(2\pi^3 / 45) g_{\tilde a} T_{\tilde a}^4 t^3$,
while $t$ = $(45/8\pi^3 g_o)^{1/2} m_{pl} T^{-2}$; the relation
between $\tilde a$ temperature and photon temperature is $T_{\tilde a}$
= $(g_o / g_{\tilde a-dg} \mu)^{1/3} T$. Accordingly, requiring
$T_{\tilde a}$ = $m_{\tilde a}$, we have that
$$
M_{D,c} \simeq \left(45 \over 128 \pi^3 \right)^{1/2}
{g_{\tilde a} g_o^{1/2} 
\over g_{\tilde a-dg}^2 \mu^2}
{m_{pl}^3 \over m_{\tilde a}^2}
\eqno(14)
$$
is $\sim 0.1 M_\odot$. Such small value of $M_{D,c}$ is due to the
actual $\tilde a$ temperature.

On the contrary
$M_{D,h} \simeq (\pi^2 / 40) t_o z_{\rm der}^{-3/2} m_{pl}^2 
\simeq 3.6 \times 10^{15} (10^4 / z_{\rm der})^{1.5} M_\odot$.
Non thermal $\tilde a$ distribution has a wide energy spread. 
Derelativization can be therefore expected to occur more gradually
than for standard 7$\, $eV neutrinos.

In the context of a phenomenologically
viable SUSY model it is therefore possible to obtain a phase space
distribution of a single particle which provides a
valid hybrid cosmological model. 

\bigskip\bigskip
\centerline{\bf Acknowledgements}
\bigskip

Thanks are due to David Lyth, Michael Dine and Joel Primack for discussions.
F.G. thanks the Departments of Physics of Milano and Padova
Universities for their kind hospitality.

\bigskip\bigskip
\centerline{\bf References}

\begin{description}
\item{1)} R.~Valdarnini and S.A.~Bonometto, Astr. $\&$
Astroph., {\bf 146}, 235 (1985); S.A.~Bonometto and R.~Valdarnini, 
Astroph.~J. {\bf 299}, L71 (1985); see also Q.~Shafi and F.W.~Stecker,
Phys. Rev. Lett. {\bf 53}, 1292 (1984) and S.~Achilli, F.~Occhionero
and R.~Scaramella, Astroph.~J. {\bf 299}, 577 (1985).

\item{2)} A.~Klypin, J.~Holtzman, J.~Primack and E. Reg\"os,
Astroph.~J. {\bf 416}, 1 (1993).

\item{3)} J.A.~Holtzman, Astroph.~J. Suppl. {\bf 71}, 1 (1989);
  A.N.~Taylor  and M.~Rowan--Robinson, Nature {\bf 359}, 396 (1992);
  J.A.~Holtzman and J.~Primack; Astrophys.~J. {\bf 396}, 113 (1992);
  a recent note by D.Yu.~Pogosyan and A.A.~Starobinski,
  CUA preprint (1993), shows that an hybrid model with $\Omega_{\rm
  HDM} \sim 0.2$, while keeping $z_{\rm der} \sim 10^4$, would provide
  an even better fit of data.

\item{4)} R.~Peccei and H.~Quinn, Phys. Rev. Lett. {\bf 38}, 1440 (1977);
Phys. Rev. {\bf D 16}, 1791 (1977).

\item{5)} M.~Dine, W.~Fischler and M.~Srednicki, Phys. Lett. {\bf 104~B},
99 (1981).

\item{6)} D.A.~Dicus, E.W.~Kolb, V.L.~Teplitz and R.V.~Wagoner, Phys.
Rev. {\bf D~22}, 839 (1980);
M.~Fukugita, S.~Watamura and M.~Yoshimura, Phys. Rev. {\bf D~26},
1840 (1981).

\item{7)} This occurred if such transition took place after a suitable
supercooling and as a deflagration in a regime next to the Jouguet curve. See 
K.~Enqvist, J.~Ignatius, K.~Kajantie and K.~Rummukainen, Phys. Rev. {\bf D~45},
3415 (1992). The amount of entropy released depends on the length of expansion
during supercooling. $T_o$ would rise to the observed value if the 
increase of the scale factor is $\sim 6$. 

\item{8)} S.A.~Bonometto, F.~Gabbiani and A.~Masiero, Phys. Lett. {\bf B~222},
433 (1989).
 
\item{9)} K.~Rajagopal, M.S.~Turner and F.~Wilczek, Nucl. Phys. {\bf B~358},
447 (1991).

\item{10)} J.F.~Nieves, Phys. Lett. {\bf B~174}, 411 (1986).
 For a previous calculation in the limit of equal top
and stop masses, see: J.E.~Kim, A.~Masiero and D.V.~Nanopoulos,
Phys. Lett. {\bf 139~B}, 346 (1984).

\item{11)} J.~Ellis {\it et al.}, Nucl. Phys. {\bf B~238}, 453 (1984).

\item{(12)} See, e.g., S.A.~Bonometto, Lett. Nuovo Cimento {\bf 1}, 677 (1971),
and S.A.~Bonometto, F.~Lucchin and P.~Marcolungo, Astr. $\&$ Astrophys.
{\bf 31}, 41 (1974).
\end{description}

\bigskip\bigskip
\centerline{\bf Figure Captions}

\begin{description}
\item{Fig. 1 --} Values of $m_{\tilde f}$ and $m_{\chi}$ leading to
$z_{\rm der} \simeq 10^4$ for $\Omega_{\rm CDM} = 4 \Omega_{\rm HDM}
$ = 0.8. Different curves (bottom to top) refer to $m_t$ from
120 GeV to 180 GeV.

\item{Fig. 2 --} $V_{\rm PQ}$ as a function of $m_{\chi}$ for $m_t$ = 140 GeV
and $z_{\rm der} \simeq 10^4$.

\item{Fig. 3 --} Evolution of densities of radiation (solid line),
CDM (dashed line) and HDM (dotted line) in logarithmic scales.
The HDM line
accounts for $\chi$'s prior to their decay and then for
non-thermal axinos. The large increase of $\rho T^{-4}$
at $T \sim$ 100 GeV is related to entropy release at the
electroweak phase transition.
\end{description}

\vfill\eject

\begin{figure}[htb]
\centering
\leavevmode
\centerline{
\epsfbox{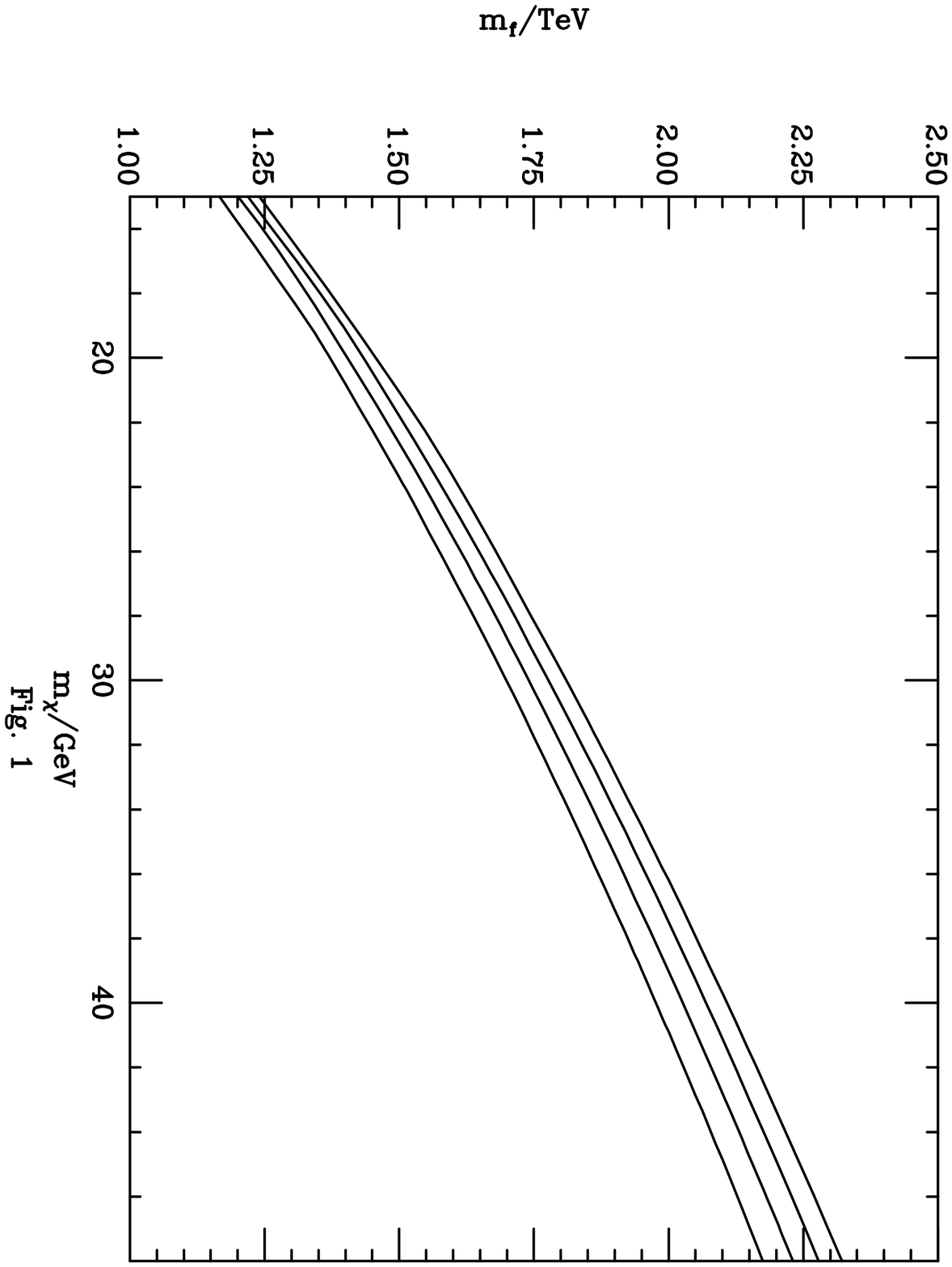}}
\end{figure}

\begin{figure}[htb]
\centering
\leavevmode
\centerline{
\epsfbox{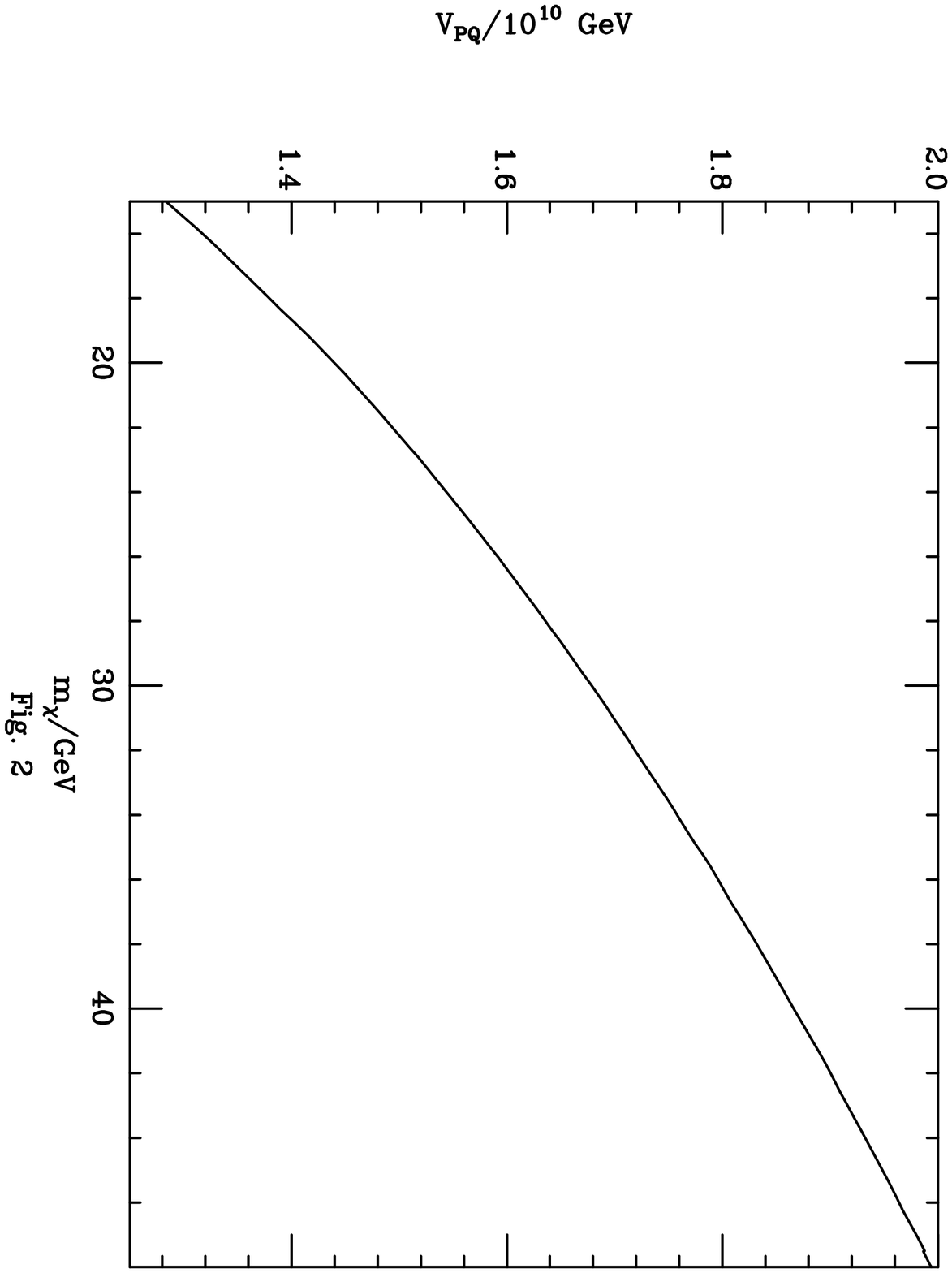}}
\end{figure}

\begin{figure}[htb]
\centering
\leavevmode
\centerline{
\epsfbox{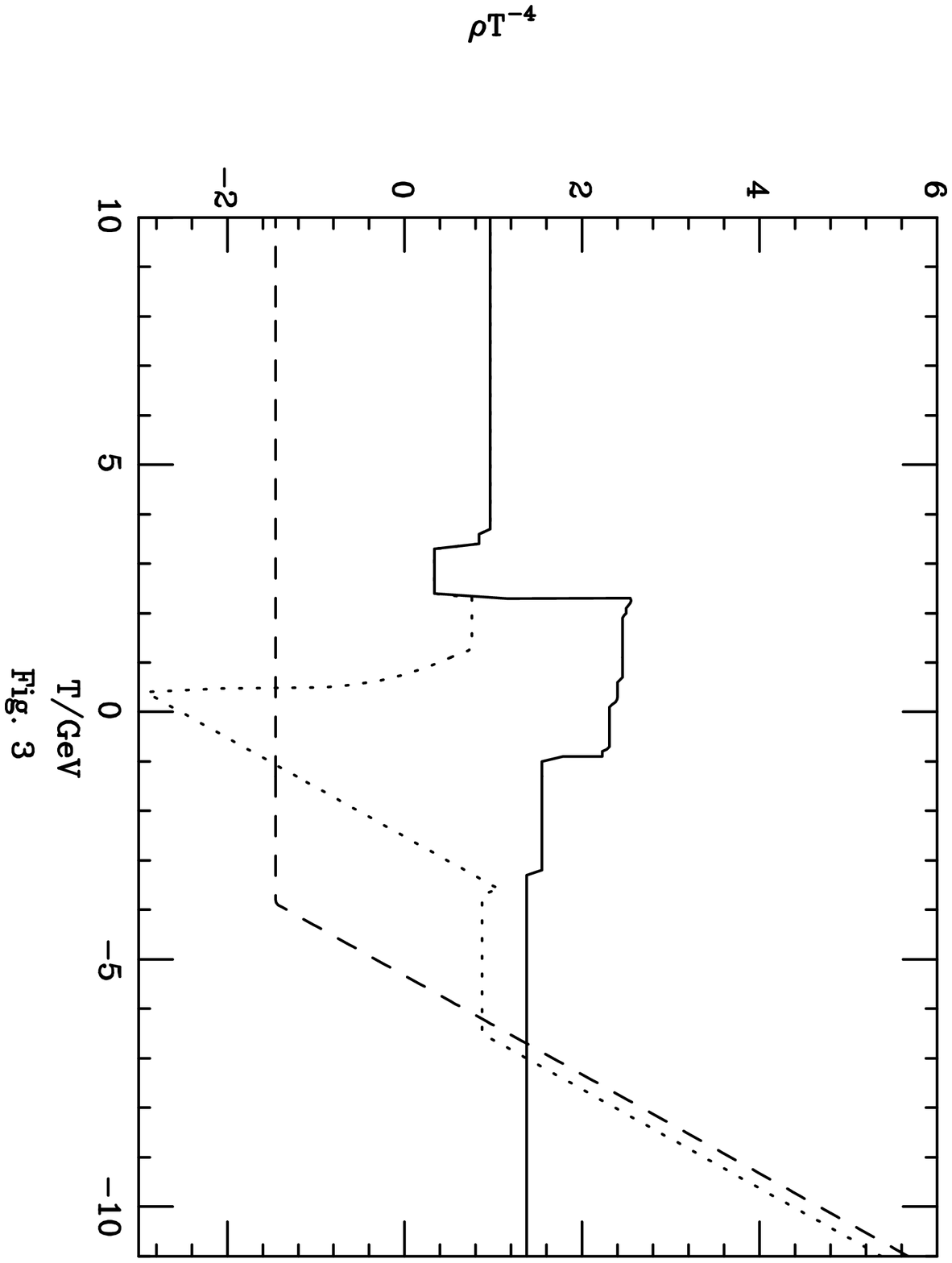}}
\end{figure}

\end{document}